\documentclass[preprint,aps]{revtex4}
\usepackage{mathrsfs}

\usepackage{graphicx}

\begin{document}

\title{Comment on arXiv:1012.1484v1 ``Structural origin of apparent Fermi surface pockets in angle-resolved photoemission of
Bi$_2$Sr$_{2-x}$La$_x$CuO$_{6+\delta}$" by King et al.}

\author{X. J. Zhou$^{1}$, Jianqiao Meng$^{1}$, Yingying Peng$^{1}$, Junfeng He$^{1}$ and Li Yu$^{1}$}

\affiliation{
\\$^{1}$National Laboratory for Superconductivity, Beijing National Laboratory for Condensed
Matter Physics, Institute of Physics, Chinese Academy of Sciences,
Beijing 100190, China
}
\date{December 16, 2010}
%
%



\maketitle


In our Nature paper\cite{Meng}, we reported observation of Fermi pocket in Bi$_2$(Sr$_{2-x}$La$_x$)CuO$_{6+\delta}$ (La-Bi2201)
by angle-resolved photoemission (ARPES) measurements.  The observed Fermi pocket is not centered around
($\pi$$\slash$2,$\pi$$\slash$2) which can unambiguously rule out its trivial origin of the ``shadow band" commonly observed in Bi2201 and Bi2212\cite{Shadow}.  Particularly, the observed Fermi pocket exhibits unusual doping dependence, i.e., it is observed in the underdoped T$_c$=18 K (x=0.73) and T$_c$=26 K (x=0.60) samples, but not in the optimally-doped T$_c$=32 K (x=0.40) and underdoped T$_c$ $\sim$3 K (x=0.84) samples.

In our Nature paper\cite{Meng}, we discussed various possible origins of the observed Fermi pocket.  (1). Additional superstructure: ``The presence of additional superstructure, which would give rise to new bands, appears to be unlikely because there is no indication of such additional bands observed in our measurements."  (2). {\it d}-density-wave: ``This particular location makes it impossible to originate from the {\it d}-density-wave ``hidden order" that gives a hole-like Fermi pocket centered around ($\pi$/2,$\pi$/2) point\cite{ChakravartyPRB,ChakravartyPNAS}." (3). Resonant valence bond picture: ``The phenomenological resonant valence bond picture\cite{FCZhang} shows a fairly good agreement with our observations, in terms of the location, shape and area of the hole-like Fermi pocket and its doping dependence.  One obvious discrepancy is that the spectral weight on the back side of the Fermi pocket near ($\pi$/2,$\pi$/2) is expected to be zero from these theories\cite{WenLee,FCZhang} which is at odds with our measurements." (4). Incommensurate density wave: ``We note that the existence of incommensurate density wave could also potentially explain the observed pockets.  One possible wave-vector needed is (1$\pm$0.092,1$\pm$0.092) which is diagonal that can be examined by neutron or X-ray scattering measurements."  (5). Charge$\slash$spin density wave: ``There are indications of the charge-density-wave (CDW) formation reported in the Bi2201 system\cite{HudsonSTM}; whether the Fermi surface reconstruction caused by such a CDW order or its related spin-density-wave order can account for our observation needs to be further explored."  We discussed pros and cons of various scenarios without pinning down on any particular possibility.

In a recent paper by King et al.\cite{King}, they reported LEED (low energy electron diffraction) and ARPES measurements on La-Bi2201 with different dopings and optimally-doped (Pb,La)-Bi2201. They reported observation of ``Fermi pocket" in their UD14.5K La-Bi2201 sample (underdoped, T$_c$=14.5 K).  They did not observe ``Fermi pocket" in their OP30K La-Bi2201 sample (optimally-doped, T$_c$=30 K).  These observations are consistent with our results and there is no disagreement between the two groups on the experimental aspect.

The key difference between King et al.'s paper\cite{King} and our paper\cite{Meng} lies in the interpretation of the data.  King et al.\cite{King} proposed that the Fermi pocket we observed is due to an additional {\bf q$_2$} superstructure.  This corresponds to the ``additional superstructure" possibility (scenario 1 in the above possibility list) we discussed in our paper\cite{Meng} that we considered ``to be unlikely".   In this Comment, we will show that King et al's interpretation has serious inconsistencies that render it highly unlikely.

Let us first examine how the Fermi surface of Bi2201 would look like when there are two co-existing {\bf q$_1$} and {\bf q$_2$} superstructures along the $\Gamma$-Y direction (Fig. 1).  In Fig. 1a, we start with a single main Fermi surface (MB, red line). In cuprate compounds, it is known that there are so-called shadow bands that could be caused by either structural or magnetic origins\cite{Shadow}.  The shadow band (SB, black dashed line) in Fig. 1a is obtained by symmetrizing  the  main band MB with respect to the ($\pi$,0)-(0,$\pi$) line, or equivalently, by shifting the  main band MB with a wavevector ($\pi$,$\pi$).

In the Bismuth-based cuprates like Bi2201, it is well-known that there is an incommensurate superstructure along the $\Gamma$-Y direction with a wavevector {\bf q$_1$}=(q$_1$,q$_1$)\cite{Umklapp}.  In general, the existence of a superstructure with a wavevector {\bf q} will give rise to replica bands from shifting the original bands by $\pm$n{\bf q}  with n being the order of the superstructure band. If there is another superstructure {\bf q$_2$}=(q$_2$,q$_2$) present, the two co-existing {\bf q$_1$} and {\bf q$_2$}  superstructures would produce replica bands from shifting the original bands by $\pm$m{\bf q$_1$}$\pm$n{\bf q$_2$} with m and n being the order of {\bf q$_1$} and {\bf q$_2$} superstructure bands, respectively. Eight first-order (m+n=1) superstructure bands are expected with co-existing {\bf q$_1$} and {\bf q$_2$} superstructures; 4 from the main band MB: MB$\pm${\bf q$_1$}, MB$\pm${\bf q$_2$}, and the other 4 from the shadow band SB: SB$\pm${\bf q$_1$}, SB$\pm${\bf q$_2$}, as shown in Fig. 1b. Sixteen second-order (m+n=2) superstructure bands can be produced with both q$_1$ and q$_2$ superstructures present;  8 from the  main band MB: MB$\pm$2{\bf q$_1$}, MB$\pm$2{\bf q$_2$}, MB$\pm$({\bf q$_1$}+{\bf q$_2$}), MB$\pm$({\bf q$_1$}-{\bf q$_2$}), and the other 8 from the shadow band SB: SB$\pm$2{\bf q$_1$}, SB$\pm$2{\bf q$_2$}, SB$\pm$({\bf q$_1$}+{\bf q$_2$}), SB$\pm$({\bf q$_1$}-{\bf q$_2$}), as shown in Fig. 1c. Much more superstructure bands will be produced if we consider the order of superstructure higher than 2 (m+n$>$2). Here we limit the order to 2 (m+n=2) because this is sufficient for our discussions.  Note that the crossings of multiple main band replicas and shadow band replicas will produce something like ``Fermi pockets". Even considering only the first-order superstructure bands, more than 4 apparent ``Fermi pockets" are already present with different shape and size (Fig. 1b).  The number of such ``Fermi pockets" increases dramatically with increasing order of the superstructure bands (Fig. 1c).

King et al. measured  LEED and ARPES on the Pb- and La- co-substituted optimally-doped (Pb,La)-Bi2201 sample (Fig. 2 in \cite{King}). From their LEED measurement(Fig. 2a in \cite{King}), they reported observation of an additional superstructure with q$_2$=0.072 $\pi$/a in addition to the usual superstructure with q$_1$=0.225 $\pi$/a .  In their ARPES measurement (Fig. 2b in \cite{King}), one can observe the first-order superstructure bands like MB$\pm${\bf q$_1$} and MB$\pm${\bf q$_2$}, some second-order superstructure bands like MB-{\bf q$_1$}+{\bf q$_2$}, shadow band, and superstructure bands of the shadow band like second-order SB-{\bf q$_1$}+{\bf q$_2$}. Overall speaking, the observed bands from their ARPES measurement are consistent with the LEED measurement and with the expected bands with two co-existing {\bf q$_1$} and {\bf q$_2$} superstructures (Fig. 1c).  King et al. marked two ``apparent pockets" in their ARPES data (Fig. 2b in \cite{King}): one corresponding to the crossing of the  main band MB and a second-order  superstructure band SB-{\bf q$_1$}+{\bf q$_2$}, the other corresponding to the crossing of the  shadow band SB and a second-order superstructure band MB+{\bf q$_1$}-{\bf q$_2$}.   As we discussed before, with two co-existing superstructures present, it is not surprising to observe such ``apparent pockets".  In fact, more than 4 ``apparent pockets" can be present when we only consider the first-order superstructures (Fig. 1b), and far more than 2 ``apparent pockets" can be expected (with different shape and size) when we consider second-order superstructures (Fig. 1c).

The addition of Pb in the (Pb,La)-Bi2201 sample apparently makes it behave quite differently from the Pb-free La-Bi2201 samples. While the optimally-doped (Pb,La)-Bi2201 sample shows an additional superstructure {\bf q$_2$} from the LEED measurement (Fig. 2a in \cite{King}) and ``apparent pockets" from the ARPES measurements (Fig. 2b in \cite{King}),  no clear indication of an additional {\bf q$_2$} superstructure is reported from the LEED pattern of the optimally-doped La-Bi2201 (Fig. 1d in \cite{King}), and no indication of ``Fermi pocket" is observed from their ARPES measurement (Fig. 1b in \cite{King}) and from our ARPES measurement\cite{Meng}.  In this sense, the (Pb,La)-Bi2201 results are not directly relevant to the issue we discuss, i.e., whether the ``Fermi pocket" observed in the underdoped La-Bi2201 can be explained by an additional superstructure. However, the (Pb,La)-Bi2201 measurement does provide a good reference for La-Bi2201 samples on how the Fermi surface would look like experimentally when there are two co-existing superstructures, as we will discuss below.

King et al. also reported observation of an additional superstructure {\bf q$_2$} with q$_2$=0.130 in UD14.5K La-Bi2201 sample in addition to the usual {\bf q$_1$} superstructure with q$_1$=0.235, as inferred from their LEED pattern (Fig. 1c in \cite{King}).  Based on this observation, they proposed that the Fermi pocket we observed can be explained by this additional superstructure {\bf q$_2$}. According to their proposal, the center Fermi pocket band LP in Fig. 1e from our vacuum ultraviolet (VUV) laser ARPES measurement is assigned as the  shadow band SB shifted by {\bf q$_2$} superstructure: SB-{\bf q$_2$};  the  pocket band LPS near the bottom-left corner in Fig. 1e is assigned as SB-{\bf q$_1$}-{\bf q$_2$},  and the  pocket band HP near the up-right corner in Fig. 1d, measured using Helium discharge lamp ARPES, is assigned as SB+{\bf q$_1$}-{\bf q$_2$}.   Although these assignments seem to provide a possible explanation for the observed pocket bands, they also create serious inconsistencies in their interpretation:\\

1). Because the proposed assignments of the pocket bands involve two second-order superstructure bands, SB-{\bf q$_1$}-{\bf q$_2$} and SB+{\bf q$_1$}-{\bf q$_2$}, in principle, all the other second-order superstructure bands,  and particularly the first-order bands,  plotted in Fig. 1c should become observable. This would give a total number of over 20 bands. In fact, in their measurements (Fig. 1a in \cite{King}), as well as in our measurements (Figs. 1d and 1e), most of these expected bands are not observed.

2). Specifically, King et al.'s proposal is hard to explain the absence of the first-order {\bf q$_2$} superstructure bands of the  main band MB, MB$\pm${\bf q$_2$}, in the ARPES measurements on the underdoped La-Bi2201.   As seen in Fig. 1a of \cite{King}, the first-order {\bf q$_1$} superstructure bands of the main band MB, MB$\pm${\bf q$_1$},  are clearly seen as expected, with obvious spectral weight below the solid green lines in Fig. 1a of \cite{King}.  With the presence of an additional {\bf q$_2$} superstructure as King et al. proposed, one would expect to see also clear first-order {\bf q$_2$} superstructure bands of the main band MB, MB$\pm${\bf q$_2$}, i.e., clear spectral weight at the location of the solid blue lines in Fig. 1a of \cite{King}. If it is true that, as they proposed, the pocket bands assigned as the second-order superstructure bands of the shadow band SB can be seen, this definitely means that the first-order MB$\pm${\bf q$_2$} bands should be observed more clearly.  But they are not observed in their measurements on the UD14.5K sample (Fig. 1a in \cite{King}); they are not seen in our Helium lamp ARPES measurement (Fig. 1d) and our high resolution VUV laser ARPES measurement (Fig. 1e).  This appears not due to the photoemission matrix element effect because the first-order MB$\pm${\bf q$_2$} bands would have the same parity as the first-order MB$\pm${\bf q$_1$} which are clearly seen (Fig. 1a of \cite{King}). Also in their (Pb,La)-Bi2201 sample (Fig. 2 in \cite{King}), these first-order q$_2$ superstructure bands of the  main band MB are indeed observed.

3). In King et al.'s proposal\cite{King}, the Fermi pocket is assigned to come from the  shadow band SB shifted by the {\bf q$_2$} superstructure.  They also tried to demonstrate that the pocket band shows similar polarization behavior as the shadow band (Fig. 3 in \cite{King}). This means that the observation of the Fermi pocket relies on the obvious presence of the shadow band SB first, and then shifting of the SB band by {\bf $q_2$} superstructure. But they do not see the  shadow band SB in their measurement on UD14.5K sample (Fig. 1a in \cite{King}) although they observed clear pocket band.  This is also the case for our measurement (Fig. 1d) where we observe the pocket band HP but we do not see the shadow band SB.  According to their proposal, how can one observe the Fermi pocket without seeing the shadow band?

In summary,  King et al. tried to assign the Fermi pocket we observed as due to an additional {\bf q$_2$} superstructure.  In the process,
it creates a number of serious inconsistencies and flaws in their interpretation as we discussed above. Any one of the above inconsistencies goes strongly against King et al.'s proposal,  making their proposed structural origin highly unlikely as the cause of the observed Fermi pocket. In our paper\cite{Meng}, we already pointed out that this structural origin is unlikely based on the absence of additional {\bf q$_2$}-induced first-order superstructure bands of the main band MB. This conclusion remains valid and gains even stronger support by considering King et al's data. Of course, to pin down the exact origin of the observed Fermi pocket,  more work needs to be done to disentangle other possibilities as we listed in our paper\cite{Meng}.

\newpage


\begin{figure*}[tbp]
\begin{center}
\includegraphics[width=0.9\columnwidth,angle=0]{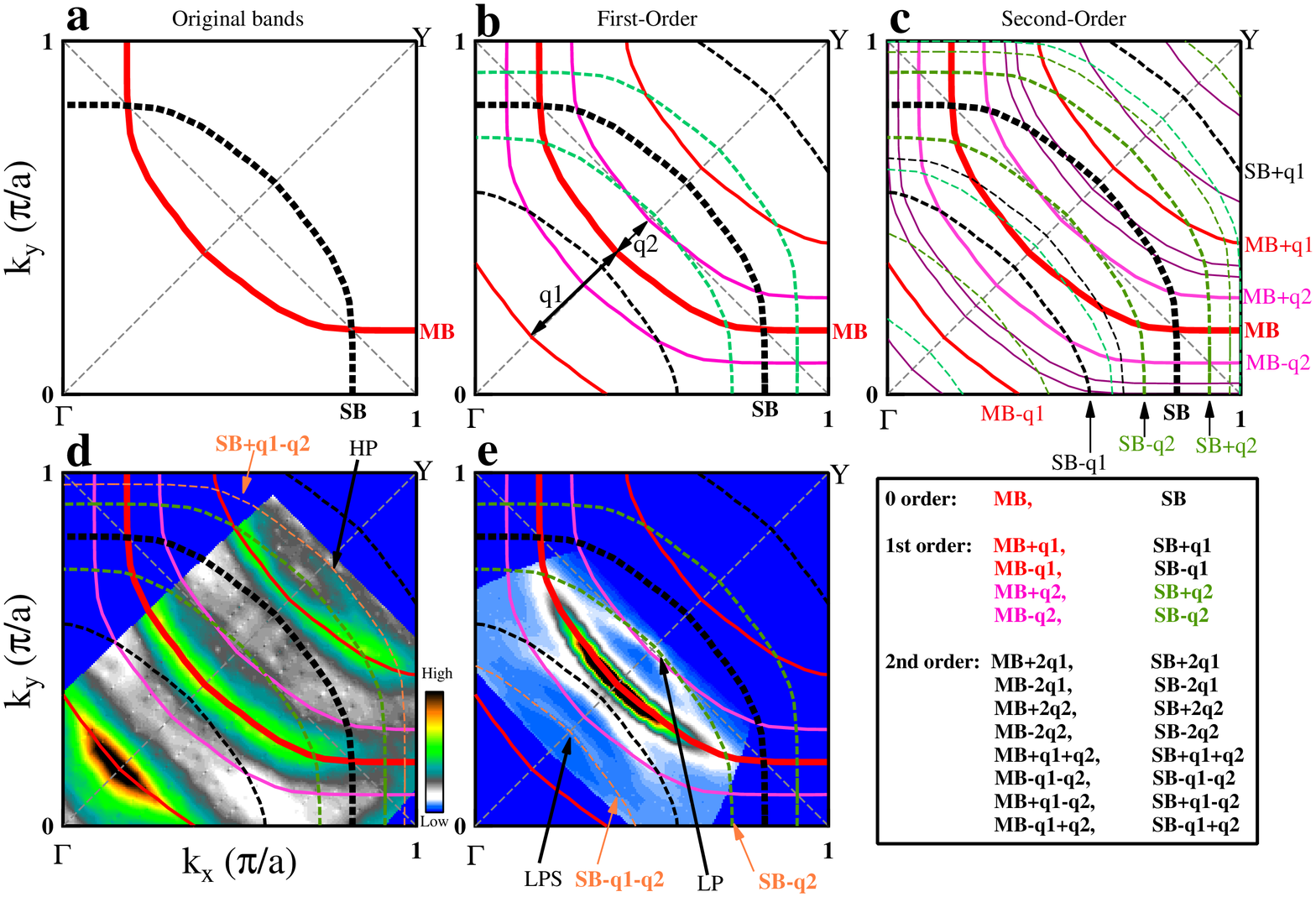}
\end{center}
\caption{Expected Fermi surface with two co-existing {\bf q$_1$} and {\bf q$_2$} superstructures and its comparison with ARPES measurements for the  La-Bi2201 UD18K sample (underdoped, x=0.73, T$_c$=18 K). (a). Main Fermi surface (MB, thick red line) and its corresponding shadow band (SB, black dashed line).  The main band MB is obtained directly from our ARPES measurement (Fig. 1d).   (b). First-order superstructure bands caused by {\bf q$_1$} and {\bf q$_2$} superstructures. Eight first-order superstructure bands are produced from shifting the original main band (MB) and shadow band (SB) by $\pm${\bf q$_1$} or $\pm${\bf q$_2$} along the $\Gamma$-Y direction (MB$\pm${\bf q$_1$}, MB$\pm${\bf q$_2$}, SB$\pm${\bf q$_1$}, SB$\pm${\bf q$_2$}).   Here q$_1$=0.24 $\pi$/a is obtained from direct ARPES measurement (Fig. 1d) while q$_2$ is 0.092 $\pi$/a chosen to make the shifted shadow band best match the pocket band, as proposed in King et al.'s paper\cite{King}. (c). Second-order superstructure bands caused by {\bf q$_1$} and {\bf q$_2$}. Sixteen second-order superstructure bands can be produced from shifting the original main band MB and shadow band SB by $\pm$2{\bf q$_1$}, $\pm$2{\bf q$_2$}, $\pm$({\bf q$_1$}+{\bf q$_2$}) or $\pm$({\bf q$_1$}-{\bf q$_2$}).  (d). Fermi surface of UD18K sample measured by Helium discharge lamp ARPES. The first-order superstructure bands are plotted for comparison. One second-order superstructure band SB+{\bf q$_1$}-{\bf q$_2$} (orange dashed line) is plotted in order to match the pocket band HP.  (e). Fermi surface of UD18K sample measured by VUV laser ARPES. The first-order superstructure bands are also plotted for comparison. One second-order superstructure band SB-{\bf q$_1$}-{\bf q$_2$} (orange dashed line) is plotted in order to match the pocket band LPS.
}
\end{figure*}


\begin{thebibliography}{99}



\bibitem{Meng} J. Q.  Meng et al., Nature {\bf 462}, 335 (2009).
\bibitem{Shadow} P. Aebi et al., Phys. Rev. Lett. {\bf 72},  2757 (1994); K. Nakayama et al.,  Phys. Rev. B {\bf 74}, 054505 (2006).
\bibitem{ChakravartyPRB} S. Chakravarty et al.,  Phys. Rev. B  {\bf 63}, 094503(2001).
\bibitem{ChakravartyPNAS} S. Chakravarty and H. Y. Kee, Proc. Natl. Acad. Sci. USA {\bf 105}, 8835(2008).
\bibitem{FCZhang} K. Y. Yang, T. M. Rice and F. C. Zhang, Phys. Rev. B. {\bf 73}, 174501 (2006); T.-K. Ng, Phys. Rev. B {\bf 71},
172509 (2005).
\bibitem{WenLee} X. G. Wen and P. A. Lee, Phys. Rev. Lett. {\bf 76}, 503 (1996).
\bibitem{HudsonSTM} W. D. Wise et al.,  Nature Phys. {\bf 4}, 696 (2008).
\bibitem{King} P. D. C. King et al., arXiv: 1012.1484.
\bibitem{Umklapp} H. Ding et al.,  Phys. Rev. Lett. {\bf 74}, 2784 (1995); J. Osterwalder et al.,  Appl. Phys. A {\bf 60}, 247 (1995).



\end{thebibliography}
\end{document}